# Magnetothermal study of Dy Stuffed Spin Ice: $Dy_2(Dy_xTi_{2-x})O_{7-x/2}$

B. G. Ueland[1*], G. C. Lau[2], R.S. Freitas[1†], J. Snyder[1‡], M. L. Dahlberg[1], B. D. Muegge[2], E. L. Duncan[2], R. J. Cava[2], and P. Schiffer[1§]

[1]*Department of Physics and Materials Research Institute, Pennsylvania State University, University Park PA 16802*

[2]*Department of Chemistry and Princeton Materials Institute, Princeton University, Princeton, NJ 08540*

## Abstract

We have studied the thermodynamics of the stuffed spin ice material, $Dy_2(Dy_xTi_{2-x})O_{7-x/2}$, in which additional $Dy^{3+}$ replace $Ti^{4+}$ in the pyrochlore $Dy_2Ti_2O_7$. Heat capacity measurements indicate that these materials lose the spin ice zero point entropy for $x \geq 0.3$, sharply contrasting with results on the analogous Ho materials. A finite ac susceptibility is observed as $T \rightarrow 0$ in both the Ho and Dy materials with $x = 0.67$, which suggests that spin fluctuations persist down to $T \sim 0$. We propose that both the entropy and susceptibility data may be explained as a result of domains of local pyrochlore type structural order in these materials.

Materials known as geometrically frustrated magnets have atomic moments which cannot simultaneously satisfy all spin-spin interactions due to their regular positions on a crystal lattice. The energy scales of the interactions in these magnets offer unique opportunities to study how frustrated thermodynamic systems settle into their lowest energy states [1,2,3]. Examples of novel ground states observed in geometrically frustrated magnets include spin-glass-like states despite the presence of minimal structural disorder [4,5,6,7], cooperative paramagnetic states, in which the spins are locally correlated yet continue to fluctuate as $T \sim 0$ [8,9,10,11], and spin ice states [12,13,14,15,16,17,18,19,20,21], in which the spins freeze into a state analogous to that of the protons in frozen water. In this paper we report experimental results for variants of two spin ice materials, formed by increasing the density of spins present in the materials.

Pure spin ices have a pyrochlore lattice with the general formula $A_2B_2O_7$, in which the $A$ sites are occupied by either of the magnetic lanthanides $Ho^{3+}$ or $Dy^{3+}$, forming a magnetic sublattice of corner sharing tetrahedra. The $B$ sites, which are typically filled by either non-magnetic $Sn^{4+}$ or $Ti^{4+}$, form a separate sublattice of corner sharing tetrahedra, which is offset from the $A$ sites by the same distance as the side of a single tetrahedron. This results in 6 $A$ nearest neighbors and 6 $B$ nearest neighbors for each $A$ or $B$ site. The local crystal field in these materials produces a highly anisotropic, Ising like single ion ground state for the rare earth ions, constraining the spins to point either towards or away from the center of each magnetic tetrahedron [22,23]. Together with the strong dipolar interactions that are present [24], the minimum energy per tetrahedron results when two spins point into and two spins point out of each tetrahedron [12,16].

This configuration has a high level of degeneracy, causing the spins to freeze into a disordered state with approximately the same zero point entropy as in water ice [17].

Recent work by our groups has demonstrated that the pyrochlore lattice can be altered by adding additional rare-earth ions to the $B$ site of the pyrochlore lattice, thus "stuffing" more moments into the system, a situation illustrated in Fig. 1 [25]. Initial studies focused on polycrystalline samples of the HoTi system, $Ho_2(Ho_xTi_{2-x})O_{7-x/2}$, $0 \leq x \leq 0.67$, dubbed "stuffed spin ice" (SSI) [25], and subsequent studies on single crystal samples of Ho SSI have been carried out by Zhou *et al* [26]. In this series of materials, fits made to low field susceptibility data using the Curie-Weiss law showed that the effective spin-spin interaction becomes more antiferromagnetic (AFM) with increasing stuffing, $x$, while, surprisingly, the residual entropy per spin remains at the spin ice ($x = 0$) value for all $x$. X-ray experiments performed to determine the crystal structure of Ho SSI found that diffraction peaks corresponding to pyrochlore order decreased in intensity with increasing $x$ for $x > 0.3$, and completely disappeared at $x = 0.67$, suggesting that Ti mixed on to the pyrochlore $A$ sites [25,27]. These data led to the conclusion that the crystal transformed into a disordered fluorite lattice, meaning that $Ho^{3+}$ and $Ti^{4+}$ were randomly distributed on both the $A$ and $B$ pyrochlore sites, presumably resulting in a disordered and frustrated mixed Ho/Ti lattice of side sharing tetrahedra [25,27].

Recent neutron diffraction and electron microscopy experiments have, however, shown that the local structure is more subtle, with features that were irresolvable in the earlier X-ray diffraction data [28]. In fact, these new data on fully stuffed Ho SSI, $Ho_{2.67}Ti_{1.33}O_{6.67}$, show that a separation of the Ho and Ti sites occurs due to the mismatch in the cation sizes, resulting in complex structural domains. In these "pyrochlore order" domains, one of the sublattices of corner-sharing tetrahedra is purely Ho while the other sublattice is a mixture of Ho/Ti. These domains, with sizes on the order of 50 Å or less in the Ho SSI [28], are defined by regions of such order, with boundaries at which the sublattices switch their chemical content between pure Ho and the Ho/Ti mixture. Because of the small domain sizes, the average structure of the lattice appears to be of the disordered fluorite type in the X-ray measurements, yet, in fact, the material locally retains a pyrochlore order (i.e. small domains exist with purely Ho on the *A* sites and Ho/Ti mixtures on the *B* sites, or vice versa).

Here we present susceptibility and heat capacity measurements on another SSI material, $Dy_2(Dy_xTi_{2-x})O_{7-x/2}$, $0 \leq x \leq 0.67$, which is based on the spin ice material $Dy_2Ti_2O_7$, and we also show low temperature susceptibility data for fully stuffed ($x = 0.67$) Ho SSI, which extends the range of our previous measurements [25]. In contrast to the Ho SSI materials, we find that the Dy SSI materials possess almost no zero point entropy for $x \geq 0.3$, which we suggest is attributable to an observed larger structural domain size.

Polycrystalline samples of $Dy_2(Dy_xTi_{2-x})O_{7-x/2}$ and $Ho_2(Ho_xTi_{2-x})O_{7-x/2}$, where $0 \leq x \leq 0.67$, were synthesized using previously published techniques [25,27]. We measured the magnetization *M* of each Dy SSI sample down to $T = 1.8$ K and in fields up to $H = 7$ T in a Quantum Design MPMS SQUID magnetometer, and made measurements of the heat capacity of each sample down to $T = 0.4$ K and in fields up to $H = 1$ T in a Quantum Design PPMS. Additionally, measurements of the ac susceptibilities, $\chi = dM/dH$, were made in fields up to $H = 9$ T using the ac susceptibility option for a Quantum Design

PPMS for $T \geq 1.8$ K and also using an inductance coil set in a dilution refrigerator to measure down to $T = 0.095$ K [29].

Figure 2a shows the values for the Weiss temperature $\theta_W$ as a function of $x$, as obtained from fits to $M(T)$ data, examples of which we show in Fig. 2b. Fits were made over $T = 10 - 20$ K, which is the same temperature range used in Ref. 25. Similar to the results for Ho SSI [25], we find that $\theta_W$ monotonically decreases with increased stuffing, indicating that the average effective interaction between spins becomes increasingly AFM as more rare earth cations are added to the lattice. Though it is true that the A and B sites in the pyrochlore lattice have differing oxygen coordination numbers, which may explain the change in $\theta_W$ with stuffing, we have not yet performed detailed local studies on the actual oxygen positions in the stuffed lattices. As such we cannot yet draw any accurate conclusions as to why $\theta_W$ is more antiferromagnetic in the stuffed lattices. In Fig. 2c, we plot the magnetization versus applied magnetic field of each sample at $T = 2$ K and see that $M(H)$ saturates for each value of $x$ near half the value expected for single $Dy^{3+}$ ions, which is consistent with previous measurements on pure spin ice [30,31,32] and measurements on Ho SSI [25]. Since the saturated magnetization is the same for each value of $x$ studied, and $M(H)$ is rather flat at high fields, these data suggest that the Dy ions have Ising like single ion ground states in Dy SSI over the whole range of stuffing. The specific heat data, presented below, reinforce this conclusion.

Figure 3 shows the specific heat $c_p$ and magnetic entropy for the Dy SSI materials. In Fig. 3a, we plot the temperature dependence of the $H = 0$ total specific heat for the $x = 0$, 0.3, and 0.67 materials. The dotted lines represent scaled specific heat data for nonmagnetic $Lu_2(Lu_xTi_{2-x})O_{7-x/2}$ with $x = 0$, 0.3, and 0.67, which were used to subtract out the lattice contributions to the specific heat. We note that this is the same procedure used to determine the lattice contributions to $c_p$ for data on Ho SSI [25]. However, contrary to the data processing procedure for the Ho SSI data, for the Dy SSI data there was no need to subtract out a low temperature peak arising from hyperfine interactions. The resulting magnetic specific heat $c_{mag}$ for each Dy SSI sample has a peak between $T = 1 - 2$ K, which broadens at higher values of $x$. Additionally, the maximum of the peak shifts to higher temperatures for $x \geq 0.3$. Figure 3b shows the magnetic entropy $S$ of each sample, obtained by integrating $c_{mag}/T$ from $T = 0.4$ K, and Fig. 3c shows $S(x)$, determined from the $S(T)$ curves at $T = 16$ K. As for the pure spin ice ($x = 0$) case, the heat capacity data suggest that there is no significant lower temperature (i.e. $T < 0.5$ K) contribution to $S$ for values of $x$ up to $x = 0.67$. The uncertainty in these measurements was determined by taking data on two pieces cut from the same original sample, for each value of stuffing, and the reproducibility in these measurements gives the maximum uncertainty in determining $S(T = 16$ K$)$ for each value of $x$ is of order 5%.

Examining Fig. 3c, we see that the $H = 0$ magnetic entropy for the Ho SSI materials (shown by the solid line) remains at the spin ice value over the whole range of stuffing [25][26]. In pure spin ices, (i.e. $Ho_2Ti_2O_7$) this value is less than the full $S = R\ln2$ expected for an Ising system by an amount corresponding to $S_{ZP} = \frac{1}{2}R\ln(3/2)$, which results from the zero point entropy associated with the frozen spin ice state. By contrast, the total magnetic entropy in Dy SSI reaches more than 90% of the expected Ising value of $S = R\ln2$ for $x \geq 0.3$. In addition, the data in Fig. 3a show that the increase in total entropy is chiefly due to the broadening of the peaks in $c_{mag}(T)$, indicating that an increase in localized correlations is likely responsible for the greater total entropy.

Furthermore, due to the amount of disorder present in the materials and taking into account the entropy results, it is hard to imagine that broader peaks in $c_p(T)$ reflect the development of long range order. We also see in Fig. 3c that the application of a magnetic field restores the entropy towards $S = $ Rln2 for $x < 0.3$, similar to results for pure spin ice [17], while the field has little effect on the entropy of the more stuffed samples, as is also seen for Ho SSI [25][26].

To investigate further the evolution of the low temperature state with stuffing, we measured the ac susceptibility of the fully stuffed Dy SSI, $Dy_{2.67}Ti_{1.33}O_{6.67}$ down to $T = 0.095$ K, and show data for the real part of the susceptibility, $\chi'$, in Fig. 4a. We find that $\chi'(T)$ displays maxima at $T = 1.7$ K and 2 K in the $f = 10$ Hz and 1 kHz data, respectively, and that $\chi'(T)$ apparently remains finite as $T \sim 0$. This is in sharp contrast to the behavior of the pure spin ices $Ho_2Ti_2O_7$ and $Dy_2Ti_2O_7$, in which $\chi'(T) = 0$ below $T = 0.5$ K as seen in Ref. 33 and 34, and shown in Fig. 4b for $Dy_2Ti_2O_7$. The suppression of the susceptibility to zero at low temperatures in pure spin ice indicates a complete freezing of the spins on the time scale of the measurements, which is consistent with magnetization data [33] and with the observed residual entropy [17,35]. The finite susceptibility as $T \sim 0$ in Dy SSI, however, indicates that some spins remain fluctuating down to the lowest temperatures measured – a somewhat curious result since stuffing in more Dy ions yields a higher density of spins and stronger average spin-spin interactions.

Figure 4d shows low temperature $f = 10$ Hz $\chi'(T)$ data for both $Dy_{2.67}Ti_{1.33}O_{6.67}$ and $Ho_{2.67}Ti_{1.33}O_{6.67}$, where the data are normalized by the magnitudes and positions of the maxima. The two data sets are strikingly similar, suggesting that the spin fluctuations persisting down to $T \sim 0$ in both materials arise from a common mechanism. These susceptibility data complement recent results from neutron spin echo measurements, which also show spin fluctuations persisting down to at least $T = 0.05$ K in $x = 0.3$ Ho SSI [26]. Although not presented here, frequency dependent susceptibility measurements made in an applied dc magnetic field show that the frequency dependence of the maximum in $\chi'(T)$ persists in static magnetic fields as high as $H = 1$ T, ruling out the material being a canonical spin glass [29].

Figure 4c shows $\chi'$ at $f = 50$ Hz as a function of an applied dc magnetic field for various temperatures. At $T = 1.8$ K, $\chi'(H)$ decreases with increasing field, reaching $\chi'(H) = 0$ for $H > 4$ T, in agreement with the $M(H)$ data shown in Fig. 2c. For lower temperatures, we observe two maxima in $\chi'(H)$, which become sharper as the temperature decreases. The lack of evidence for long range order in $Dy_{2.67}Ti_{1.33}O_{6.67}$ in zero field, as seen from the $\chi'(T)$ data, suggests that maxima in $\chi'(H)$ are due to the applied field breaking localized spin-spin correlations. Curiously, the field dependence of the susceptibility of $Dy_{2.67}Ti_{1.33}O_{6.67}$ is reminiscent of that for the spin liquid $Gd_3Ga_5O_{12}$ (GGG), showing a broad peak at higher temperatures, which was found to correspond to the quenching of local antiferromagnetic correlations [7]. The two peaks which evolve at lower temperatures are also reminiscent of the (much sharper) peaks associated with field-induced long range order in that material [9].

Since the susceptibility is so similar in the Ho and Dy SSI materials, the question arises of why the entropy is so different, i.e., why there is no measurable macroscopic zero point entropy in the Dy material. A possible explanation is that the dissimilarity in the entropy data arises from the differences in the structural domains of local pyrochlore order. Contrary to data for Ho SSI, X-ray diffraction data for Dy SSI do show peaks corresponding to pyrochlore order for all values of $x$ [27], and recent electron microscopy measurements, similar to those performed on Ho SSI [28], show that the pyrochlore domains in Dy SSI are an order of magnitude larger than those in the Ho SSI [36]. We hypothesize that the observed zero point entropy in Ho SSI may be associated with spins in the boundary regions between such domains, and the larger domains found in Dy SSI result in the observation of less zero point entropy due to the relatively smaller fraction of spins in such boundary regions. More structural order has been seen in $Ho_{2.67}Ti_{1.33}O_{6.67}$ samples grown in a floating zone furnace when compared to samples made in a more conventional manner [28], and the measured entropy was found to be almost 10% higher than that found in conventionally synthesized Ho SSI samples -- indicating a lower zero point entropy (although the 10% effect is near the level of uncertainty in the measured entropy for Ho SSI) [25]. Under this interpretation, the susceptibility, which is very similar for both materials, is apparently probing the spins within the bulk of the pyrochlore order domains. This explanation for our susceptibility results is similar to the one put forth by Zhou *et al* [26], in which they claim separated, unfrozen regions of spins as being responsible for the finite susceptibility and constant entropy with stuffing seen in Ho SSI. Zhou *et al* suggest that the majority of frozen spins in Ho SSI are still in an ice like state, which is why the entropy remains around the spin ice value [26]. Our new data on Dy SSI further suggest that the smaller structural domains could ultimately be responsible for the measured entropy in the Ho SSI.

In summary, we have studied the Dy stuffed spin ice material and shown that it has very similar susceptibility to the analogous Ho material, but does not display the same zero point entropy. The contrasting behavior of these two material systems provide an example of the complex consequences of introducing disorder to a frustrated magnet where subtle structural differences may lead to important differences in the thermodynamic properties.

**ACKNOWLEDGEMENTS**

We gratefully acknowledge helpful discussions and exchanges with R. Moessner, M. J. P. Gingras, X. Ke and support from NSF grant DMR-0353610 and DMR-0701582. R.S.F. thanks CNPq-Brazil for sponsorship.

**Figure Captions:**

FIG. 1 A schematic representation of stuffing the pyrochlore lattice. The solid circles represent $Dy^{3+}$ cations, and the open circles represent $Ti^{4+}$. The upper drawing shows the $x = 0$ case of separate corner sharing tetrahedra solely consisting of either Dy or Ti. The bottom drawing shows the situation after an additional Dy replaces a Ti.

FIG. 2 Magnetization data for $Dy_2(Ti_{2-x}Dy_x)O_{7-x/2}$, $0 \leq x \leq 0.67$. (a) The Weiss Temperature $\theta_W$ as a function of stuffing $x$ obtained from fits to data taken in a field of $H = 0.1$ T; examples of the fits are shown in (b). (c) $M(H)$ at $T = 2$ K for each value of $x$.

FIG. 3 (a) The specific heat versus temperature of $Dy_2(Ti_{2-x}Dy_x)O_{7-x/2}$, $x = 0, 0.3, 0.67$, at $H = 0$. The dashed lines represent scaled data on non-magnetic $Lu_2(Ti_{2-x}Lu_x)O_{7-x/2}$, $x = 0$, 0.3, 0.67, which were used to subtract out the lattice contributions to $c_p(T)$. (b) The $H = 0$ magnetic entropy obtained by integrating the magnetic specific heat from below $T = 0.5$ K. The lower and upper solid lines represent the total entropy for a spin ice like and Ising like magnet, respectively. (c) The total magnetic entropy versus stuffing at $H = 0$ and 1 T. The solid line shows the total entropy versus $x$ for $Ho_2(Ti_{2-x}Ho_x)O_{7-x/2}$ at $H = 0$ [25].

FIG. 4 (a)(b) The real part of the susceptibility versus temperature at $H = 0$ of $Dy_{2.67}Ti_{1.33}O_{6.67}$ (a) and $Dy_2Ti_2O_7$ (b). (c) The $f = 50$ Hz real part of the susceptibility versus applied DC field of $Dy_{2.67}Ti_{1.33}O_{6.67}$, at various temperatures taken at a sweep rate of 0.001T/min. (d) $\chi'(T)$ at $f = 10$ Hz for $Dy_{2.67}Ti_{1.33}O_{6.67}$ and $Ho_{2.67}Ti_{1.33}O_{6.67}$ normalized by the magnitude and position of the maxima. Note that the data for $Dy_{2.67}Ti_{1.33}O_{6.67}$, deviates at low temperatures, remaining finite as $T \sim 0$.

FIG. 1 Ueland *et al.*

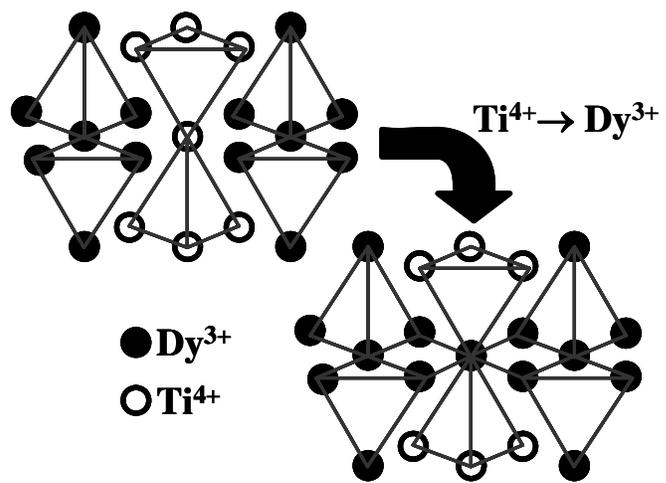

FIG. 2 Ueland *et al.*

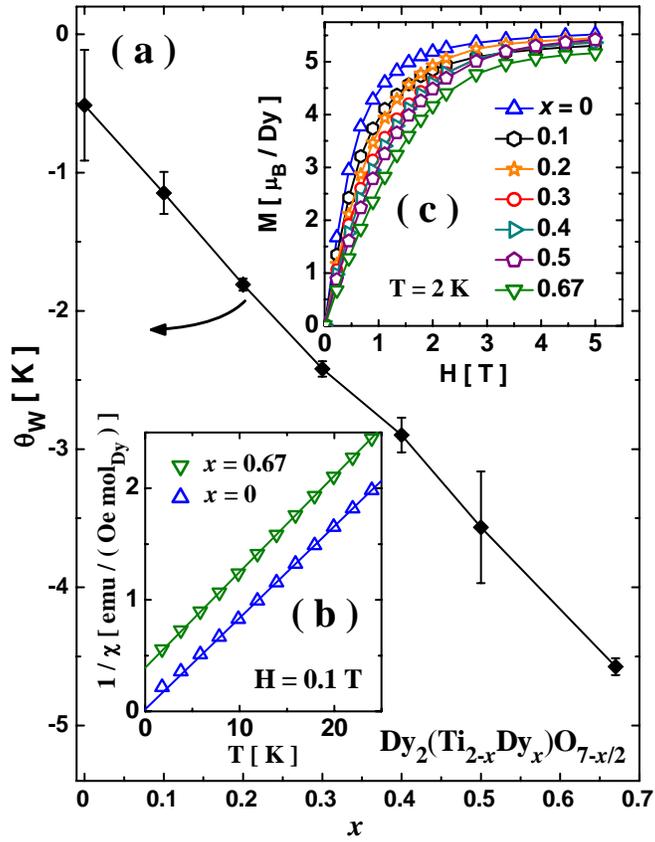

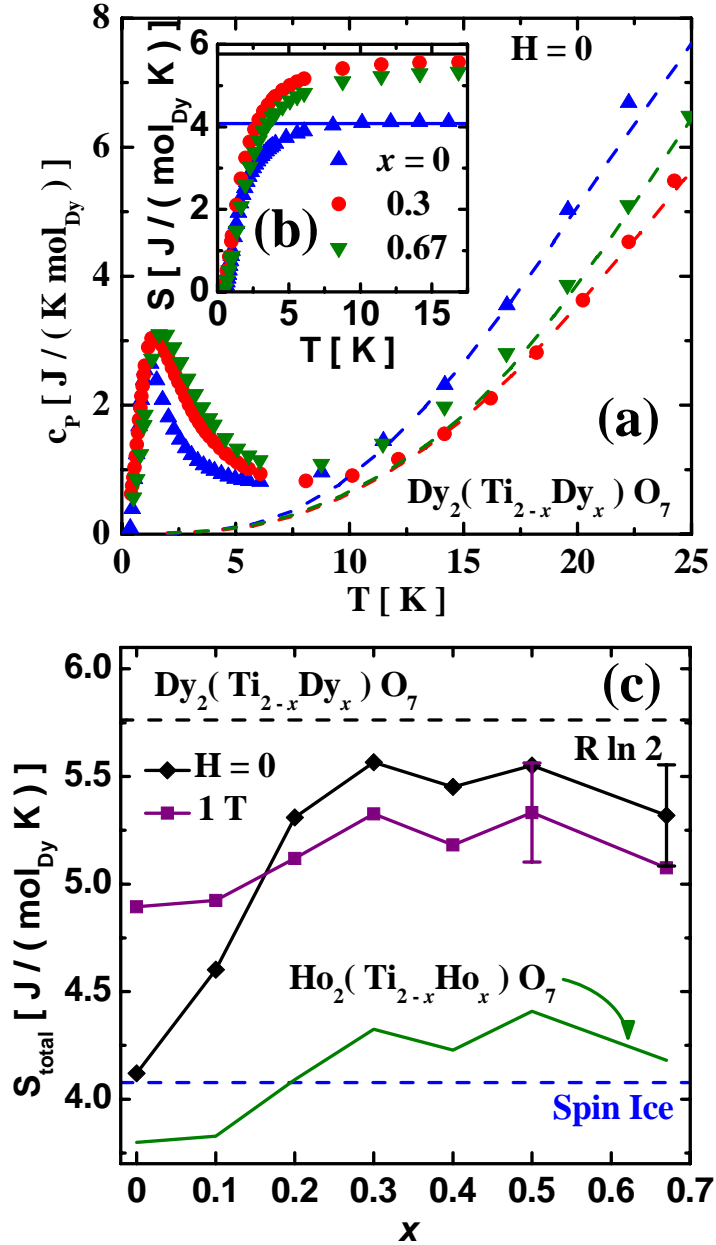

FIG. 4 Ueland *et al.*

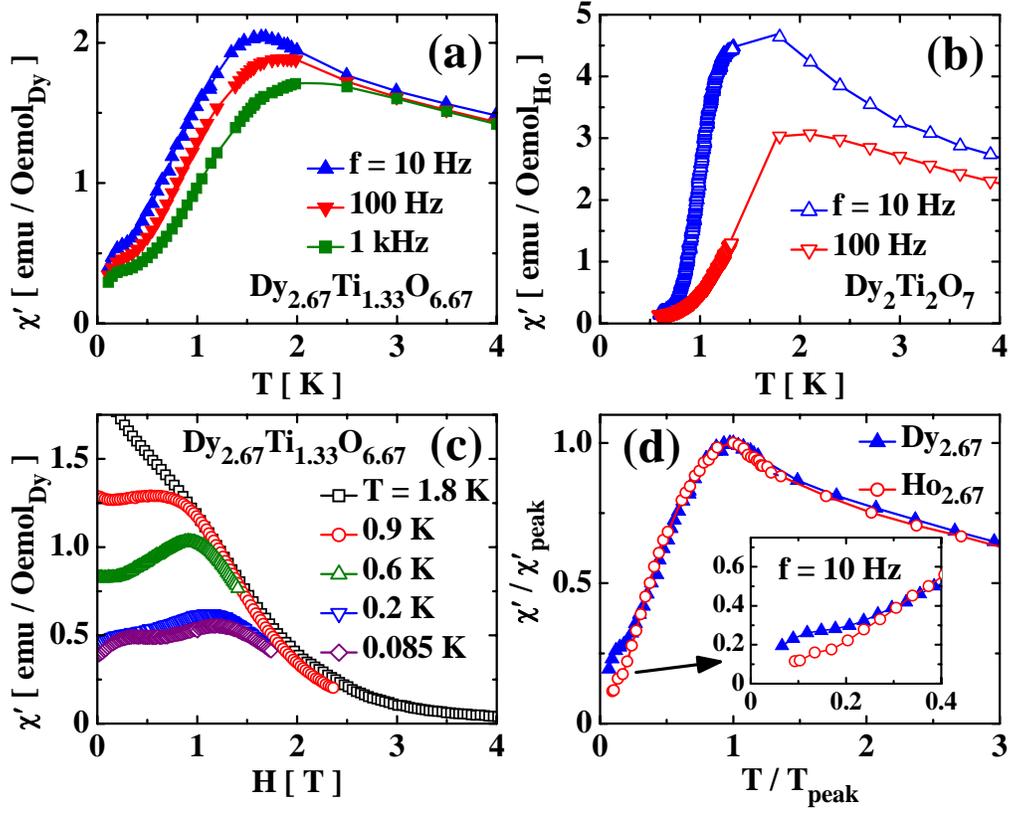

[*] Present address: NIST Center for Neutron Research, National Institute if Standards and Technology, Gaithersburg, MD 20899.

[†] Present address: Instituto de Física, Universidade de São Paulo, Caixa Postal 66318, 05315-970, São Paulo, SP, Brazil.

[‡] Present address: Scientific Instruments, Inc., West Palm Beach, FL 33407.

[§] Corresponding author: pes12@psu.edu.